\documentclass[aps, preprint]{revtex4}

\usepackage{amsfonts}
\usepackage{amssymb,epsfig}
\usepackage{latexsym}
\usepackage{amsmath}
\usepackage{graphicx}

\begin{document}

\title{Interacting entropy-corrected holographic scalar field models in non-flat universe}
\author{A. Khodam-Mohammadi\footnote{%
E-mail: \texttt{khodam@basu.ac.ir}}~~and\ M. Malekjani\footnote{%
E-mail: \texttt{malekjani@basu.ac.ir}}}

\address{Physics Department, Faculty of Science, Bu-Ali Sina
University, Hamedan 65178, Iran}

\begin{abstract}
In this work we establish a correspondence between the tachyon,
K-essence and dilaton scalar field models with the interacting
entropy-corrected holographic dark (ECHD) model in non-flat FRW
universe. The reconstruction of potentials
 and dynamics of these scalar fields according to the evolutionary
behavior of the interacting ECHDE model are be done. It has been
shown that the phantom divide can not be crossed in ECHDE tachyon
model while it is achieved for ECHDE K-essence and ECHDE dilaton
scenarios. At last we calculate the limiting case of interacting
ECHDE model, without entropy-correction.
\end{abstract}

\maketitle


\section{Introduction \label{intro}}

The accelerated expansion of the universe is a fact which is proved
by numerous cosmological observations of supernova type Ia (SNIa),
cosmic microwave background (CMB) anisotropy and large scale
structure (LSS). Up to now scientist believe that a component named
 dark energy (DE), which possesses negative pressure, is the source
of this expansion. On the other hand, a curvature driven
acceleration model which is called, modified gravity, has been
proposed  by Strobinsky \cite{Stra} and Kerner \cite{Ker} et al.,
for the first time, in 1980. Modified gravity approach suggests the
gravitational alternative for unified description of inflation, dark
energy and dark matter with no need of the hand insertion of extra
dark components \cite{Ody1}. Among the DE scenarios, the
cosmological constant, $\Lambda$, as vacuum energy density, with the
equation of state $w_d=-1$ is the most well known theoretical
candidate of DE. However, it suffers the so-called "fine-tuning" and
"cosmic coincidence" problems \cite{Copeland}. The former asks why
the vacuum energy density is so small \cite{Weinberg} and the latter
says why vacuum energy and dark matter nearly equal today. In order
to solve these two problems, several models of dynamical dark energy
models, whose equation of state is no longer a constant but evolves
with time, have been proposed. The dynamical dark energy can be
realized by scalar fields. Scalar field models arise in string
theory and are studied as a candidates of dark energy. It includes
quintessence \cite{Wetterich}, K-essence \cite{Chiba}, phantoms
\cite{Caldwell1}, tachyon \cite{Sen}, dilaton \cite{Gasperini},
quintom \cite{Elizalde1} and so forth. The other models including
Chaplygin gas \cite{Kamenshchik}, braneworld \cite{Deffayet},
interacting holographic DE (HDE) \cite{coh99, hora, li04} and
interacting agegraphic DE (ADE) models \cite{Cai} have also been
investigated.

Among of above models, the HDE model based on holographic principle
attracted a great deal of attention in the last decade. According to
the holographic principle, the number of degrees of freedom in a
bounded system should be finite and has a relationship with the area
of its boundary \cite{Hooft}. The holographic principle is a
fundamental principle in quantum gravity. In quantum field theory, a
short distance (UV) cut-off, $\Lambda $, is related to the long
distance (IR) cut-off, $L$, due to the limit set by forming a black
hole. In the other words, the zero-point energy of a system with
size $L$ should not exceed the mass of a black hole with the same
size. This fact directs us to $L^{3}\Lambda ^{3}\leq (M_{p}L)^{3/2}$
\cite{8,coh99}. From this inequality, one can obtain a limit for
energy density corresponding to the zero point energy and cut off
$\Lambda$ as $\rho _{\Lambda }\leq M_{p}^{2}L^{-2}$ or $\rho
_{\Lambda }=3n^{2}M_{p}^{2}L^{-2}$, where $\rho _{\Lambda }\backsim
\Lambda ^{4}$. Here $n$ is a numerical constant and coefficient $3$
is given for convenience. The numerical constant $n$ has been
constrained by recent observation \cite{li1}. Applying the
holographic principle to cosmology, the HDE as an interacting
dynamical dark energy model has been constructed \cite{li04}. If we
take $L$ as a size of current universe or particle horizon, the
accelerated expansion of the universe can not be derived by HDE
model\cite{hsu}. However, in the case of future event horizon as a
length scale, HDE model can derive the universe with accelerated
expansion \cite{li04}. The holographic dark energy has been studied
widely in the literature \cite{holog}. In the HDE scenario, the
energy density depends on the entropy-area relationship of black
holes in Einstein gravity. This entropy is given as $S_{\mathrm{BH}}
= A/(4G)$, where $A\sim L^2$ is the area of horizon. However in the
context of loop quantum gravity (LQG) \cite{LQG}, this entropy-area
relationship can be modified from the inclusion of quantum effects.
The quantum corrections provided to the entropy-area relationship
leads to the curvature correction in the Einstein-Hilbert action and
vice versa \cite{zhu, Ody3}. The corrected entropy is given by
\cite{modak}
\begin{equation}
S_{\mathrm{BH}}=\frac{A}{4G}+\tilde{\alpha}\ln{\frac{A}{4G}}+\tilde{\beta},
\label{MEAR}
\end{equation}
where $\tilde{\alpha}$ and $\tilde{\beta}$ are dimensionless
constants of order unity. Considering the entropy correction, the
energy density of entropy-corrected holographic dark energy (ECHDE)
can be given as \cite{wei1}
\begin{equation}  \label{rhoS}
\rho _{\Lambda}=3n^2M_{p}^{2}L^{-2}+\alpha L^{-4}\ln
(M_{p}^{2}L^{2})+\beta L^{-4}.
\end{equation}
The first term on right hand side is the usual holographic energy
density and the other terms are due to entropy correction. The
corrections make sense only at the early stage of the universe
because the last two terms in Eq. (\ref{rhoS}) can be comparable to
the first term only when L is very small. When the universe becomes
large, ECHDE reduces to the ordinary HDE.\newline Recently, the
holographic and agegraphic dark energy models have been extended
regarding to the entropy correction and a thermodynamical
description of the ECHDE model in a universe with spacial curvature
has been done \cite{Sheykhi5}. Also the correspondence of these
models with some scalar fields and Chaplygin gas model have been
performed \cite{jamil}. Also, nowadays, many authors are interested
to consider non-flat FRW universe. The tendency of a closed universe
is shown in a suite CMB experiments \cite{Sie}. Besides of it, the
measurements of the cubic correction to the luminosity-distance of
supernova measurements reveal a closed universe \cite{Caldwell}. In
accordance of all mentioned above, we prefer to consider a model
including interacting dark matter and dark energy with a non-flat
universe. We know that the scalar field models of dark energy are
the effective theories of an underlying theory of dark energy. The
fundamental theories including supersymmetric field theories and
string/M theory can not predict the potential of scalar field
uniquely. Consequently, it is meaningful to reconstruct the
potential of DE model so that these scalar fields can describe the
evolutionary behavior of the quantum gravity theory such as HDE
model.
\newline In this work we present a correspondence between
the interacting entropy-corrected holographic dark energy (ECHDE)
and the scalar filed models: K-essence, tachyon and dilaton in
non-flat FRW universe. We extend the interacting ECHDE model by
reconstruction the potentials and the dynamics of these scalar field
models. The paper is organized as follows: In section \ref{sec2},
the EoS parameter of interacting ECHDE in FRW universe with spatial
curvature is calculated. In sections \ref{sec3}, \ref{sec4} and
\ref{sec5}, the correspondence between the above mentioned scalar
fields with interacting ECHDE in non-flat universe and
reconstruction of the potentials and dynamics are presented. We
concluded this work in section \ref{sec6}.

\section{Interacting ECHDE model in a non-flat universe \label{sec2}}

Following \cite{faroog}, we introduce the interacting ECHDE model in
non flat FRW universe. The FRW metric for non flat universe is given
by
\begin{equation}
ds^{2}=-dt^{2}+a^{2}(t)\left[ \frac{dr^{2}}{1-kr^{2}}+r^{2}(d\theta
^{2}+\sin ^{2}\theta d\varphi ^{2})\right],
\end{equation}
where $a(t)$ is the dimensionless scale factor and $k=-1, 0, 1$
represent the open, flat and closed universe, respectively. For the
non flat FRW universe, the Friedmann equation can be written as
\begin{equation}  \label{frid1}
H^{2}+\frac{k}{a^{2}}=\frac{1}{3M_{p}^{2}}[\rho _{\Lambda }+\rho
_{m}],
\end{equation}
where $M_p=(8\pi G)^{-1/2}$ is modified Planck mass, $H$ is the
Hubble parameter and $\rho_{\Lambda}$ and $\rho_{m}$ are the energy
densities of dark energy and dark matter, respectively. With the
definition of dimensionless parameters $\Omega _{m}=\rho _{m}/\rho
_{c}$, $\Omega _{\Lambda }=\rho _{\Lambda }/\rho _{c}$ and $\Omega
_{k}=k/(aH)^{2}$, corresponding to energy density of matter, dark
energy and curvature, respectively, the Friedmann equation (
\ref{frid1}) is written as
\begin{equation}
1+\Omega _{k}=\Omega _{\Lambda }+\Omega _{m}.
\end{equation}%
Here $\rho_c=3H^2M_p^2$ is the critical density. By assuming an
interaction between dark matter and dark energy, the energy
conservation equation is separated for two energy density component,
of matter and dark energy as
\begin{eqnarray}
&&\dot{\rho}_{\Lambda}+3H\rho_{\Lambda} (1+w_{\Lambda}) =-Q,  \label{quent} \\
&&\dot{\rho}_{m}+3H\rho _{m} =Q,
\end{eqnarray}%
where $w_{\Lambda}$ is EOS parameter of DE. The term
$Q=3b^2H(\rho_m+\rho_{\Lambda})$ is an energy exchange term and
$b^2$ is a coupling parameter between dark matter and dark energy.
However the nature of dark matter and dark energy remains unknown,
different lagrangian have been proposed to generate this interaction
term \cite{Tsuji}. Also above expression for $Q$ may look purely
phenomenologically. It is worthwhile to mention that this
phenomenological description of interaction may be proved with some
recent observations in galaxy clusters \cite{Bertolami}, and SNIa,
CMB, LSS, and age constraints \cite{Wang}. Three forms of
interaction has been proposed by authors. Also the dynamic of
interacting dark energy models have been studied. At first this
expression was used in the study of the suitable coupling between a
quintessence scalar field and a pressureless cold dark matter
\cite{Amendola}.

The energy density of ECHDE (\ref{rhoS}) contains a length scale $L$
as a future event horizon. Since we have
\begin{equation}  \label{L}
L=a(t)r(t),
\end{equation}
where $a(t)$ is scale factor and $r(t)$ is relevant to the future
event horizon of the observable universe. The size of future event
horizon can be defined as
\begin{equation}
R_{h}=a(t)\int\limits_{t}^{\infty }\frac{dt^{\prime }}{a(t^{\prime })}%
=a(t)\int\limits_{0}^{r_{1}}\frac{dr}{\sqrt{1-kr^{2}}}.
\label{FEscale}
\end{equation}
Giving the fact that
\begin{equation}
\int\limits_{0}^{r_{1}}\frac{dr}{\sqrt{1-kr^{2}}}=\frac{1}{\sqrt{|k|}}\text{%
sinn}^{-1}(\sqrt{|k|}r_{1})=\begin{cases}\text{sin}^{-1}(r_1) , & \,
\,k=+1,\\ r_1, & \, \, k=0,\\ \text{sinh}^{-1}(r_1), & \, \,k=-1,\\
\end{cases}
\end{equation}%
we can easily derive
\begin{equation}  \label{L1}
L=a(t)\frac{\text{sinn}(\sqrt{|k|}y)}{\sqrt{|k|}},\ \ \
y=\frac{R_{h}}{a(t)},
\end{equation}%
Considering the energy density of ECHDE ({\ref{rhoS}), the following
relation can be obtained
\begin{equation}  \label{L2}
HL=\sqrt{\frac{3n^{2}m_{p}^{2}+\alpha L^{-2}\ln (m_{p}^{2}L^{2})+\beta L^{-2}%
}{3m_{p}^{2}\Omega _{\Lambda }}}.
\end{equation}%
By differentiating Eq.(\ref{L1}) with respect to time, using Eq.(\ref{L2}%
) and the fact that $\dot{y}=a$ from Eq. (\ref{FEscale}), we obtain
\begin{equation}  \label{ldot}
\dot{L}=\sqrt{\frac{3n^{2}m_{p}^{2}+\alpha L^{-2}\ln
(m_{p}^{2}L^{2})+\beta L^{-2}}{3m_{p}^{2}\Omega _{\Lambda
}}}-\text{cosn}(\sqrt{|k|}y),
\end{equation}%
where $cosn(x)$ is the derivative of $sinn(x)$ and we also have
\begin{equation}
\frac{1}{\sqrt{|k|}}cosn(\sqrt{|k|}y)=\begin{cases}cos(\sqrt{|k|}y)
, & \,
\,k=+1,\\1, & \, \, k=0,\\ cosh(\sqrt{|k|}y), & \, \,k=-1.\\
\end{cases}
\end{equation}
Taking derivative of Eq.(\ref{rhoS}) with respect to time and using Eq.(\ref%
{ldot}), we get
\begin{eqnarray}  \label{rodot}
\dot{\rho}_{\Lambda } &=&(2\alpha L^{-5}-4\alpha L^{-5}\ln
(M_{p}^{2}L^{2})-4\beta L^{-5}-6n^{2}M_{p}^{2}L^{-3})  \nonumber \\
&&\times \left[ \sqrt{\frac{3n^{2}M_{p}^{2}+\alpha L^{-2}\ln
(M_{p}^{2}L^{2})+\beta L^{-2}}{3M_{p}^{2}\Omega _{\Lambda }}}-\text{cosn}(%
\sqrt{|k|}y)\right] .
\end{eqnarray}%
Inserting Eq. (\ref{rodot}) in Eq. (\ref{quent}), the EoS parameter
of interacting ECHDE model can be obtained as

\begin{eqnarray}  \label{eos_ec}
w_{\Lambda } &=&-1-\frac{2(-2D_L+3n^2M_p^2L^2+\alpha)\left(1-\sqrt{\frac{%
3M_p^2L^2\Omega_{\Lambda}}{D_L}}\text{cosn}(\sqrt{|k|}y)\right)}{3D_{L}}-%
\frac{b^{2}(1+\Omega _{k})}{\Omega _{\Lambda}},
\end{eqnarray}%
where we define the parameter $D_L$ as
\begin{equation}
D_L=3n^2M_p^2L^2+\alpha\ln{M_p^2L^2} +\beta=L^4\rho_{\Lambda}.
\end{equation}
As we can see from (\ref{eos_ec}), the phantom divide ($w_{\Lambda
}<-1$) may cross for a wide range of parameters $n, b, \alpha$ and
$\beta$. Setting $\alpha=\beta=0$ in Eq.(\ref{eos_ec}), we have
$D_L=3n^2M_p^2L^2$, and the EoS parameter of interacting holographic
dark energy in non flat universe can easily obtained as
\begin{equation}  \label{eos_HDE}
w_{\Lambda}=-\frac{1}{3}-\frac{2}{3}\frac{\sqrt{\Omega_{\Lambda}}}{n}\text{cosn}(%
\sqrt{|k|}y)-\frac{b^{2}(1+\Omega _{k})}{\Omega _{\Lambda}}.
\end{equation}
In the case of HDE without interaction, $b=0$, the last term of
Eq.({\ref{eos_HDE}) will be zero which is the same as Rel. (23) of
Ref. \cite{setare2}.
\section{Interacting ECHDE tachyon model \label{sec3}}

Tachyon is an unstable field which has been described in string
theory through its role in the Dirac-Born-Infeld (DBI) action to
describe the D-bran action \cite{7}. The effective Lagrangian for
the tachyon field is given by
\[
\mathcal{L}=-V(\phi )\sqrt{1-g^{\mu \nu }\partial _{\mu }\phi \partial _{\nu
}\phi },
\]%
where $V(\phi )$ is the potential of tachyon. The energy density and
pressure for the tachyon field are as following \cite{7}
\begin{equation}
\rho _{\phi }=\frac{V(\phi )}{\sqrt{1-\dot{\phi}^{2}}},  \label{tach1}
\end{equation}%
\begin{equation}
p_{\phi }=-V(\phi )\sqrt{1-\dot{\phi}^{2}}.
\end{equation}%
The EoS parameter of tachyon can be obtained as
\begin{equation}
w_{\phi }=\frac{p_{\phi }}{\rho _{\phi }}=\dot{\phi}^{2}-1.  \label{eos_tach}
\end{equation}%
Now we can establish the correspondence between the interacting ECHDE
scenario and the tachyon scalar field model. By comparing Eqs.(\ref{rhoS}) and (\ref%
{tach1}), we have
\begin{equation}
V(\phi )=\frac{D_{L}}{L^{4}}\sqrt{1-\dot{\phi}^{2}} \label{vphi},
\end{equation}
while equating Eqs.(\ref{eos_ec}) and (\ref{eos_tach}), give us
$\dot{\phi}$ as
\begin{equation}
\dot{\phi}^2=1+w_{\Lambda}=\frac{2(2D_{L}-3n^{2}M_{p}^{2}L^{2}-\alpha )\left( 1-\sqrt{%
\frac{3M_{p}^{2}L^{2}\Omega _{\Lambda }}{D_{L}}}\text{cosn}(\sqrt{|k|}%
y)\right) }{3D_{L}}-\frac{b^{2}(1+\Omega _{k})}{\Omega _{\Lambda }}.
\label{phitach}
\end{equation}%
Using the Eqs.(\ref{vphi}) and (\ref{phitach}), we can write the
potential of tachyon as
\begin{eqnarray}
V(\phi )&=&\frac{D_{L}}{L^{4}}\sqrt{-w_{\Lambda}}=\label{Vtach} \\
&&\frac{D_{L}}{L^{4}}\sqrt{1-\frac{2(2D_{L}-3n^{2}M_{p}^{2}L^{2}-%
\alpha )\left( 1-\sqrt{\frac{3M_{p}^{2}L^{2}\Omega _{\Lambda }}{D_{L}}}\text{%
cosn}(\sqrt{|k|}y)\right) }{3D_{L}}+\frac{b^{2}(1+\Omega
_{k})}{\Omega _{\Lambda }}}.\nonumber
\end{eqnarray}%
As we can see from Eqs.(\ref{phitach}) and (\ref{Vtach}), the
kinetic energy $\dot{\phi}^2$ and potential $V(\phi )$ may exist
provided that
\begin{equation}
-1\leq w_{\Lambda}\leq 0.
\end{equation}
This condition implies that the phantom divide can not be crossed in
a universe with an accelerated expansion.\newline In the limiting
case of $\alpha =\beta =0$, the kinetic energy and potential for
usual holographic tachyon dark energy without the interaction term
($b=0$) can be obtained exactly the same as ones in Ref.
\cite{setare2} as:
\begin{equation}
\dot{\phi}=\sqrt{\frac{2}{3}}\sqrt{1-\frac{\sqrt{\Omega_{\Lambda}}}{n}%
\text{cosn}(\sqrt{|k|}y))},
\end{equation}%
\begin{equation}
V(\phi
)=\frac{\sqrt{3}n^{2}M_{p}^{2}}{L^2}\sqrt{1+2\frac{\sqrt{\Omega_{\Lambda}}}{n}\text{cosn}(\sqrt{|k|}y)}.
\end{equation}%
Using $\dot{\phi}=\phi ^{\prime }H$ and Eq.(\ref{phitach}), we get
\begin{equation}
\phi ^{\prime }=\frac{1}{H}\sqrt{\frac{2(2D_{L}-3n^{2}M_{p}^{2}L^{2}-\alpha
)\left( 1-\sqrt{\frac{3M_{p}^{2}L^{2}\Omega _{\Lambda }}{D_{L}}}\text{cosn}(%
\sqrt{|k|}y)\right) }{3D_{L}}-\frac{b^{2}(1+\Omega _{k})}{\Omega _{\Lambda }}%
}  \label{phiprim1}
\end{equation}%
Hence, we can find the evolutionary form of the tachyon scalar field
as follows
\begin{eqnarray}
&&\phi (a)-\phi (a_{0})=  \label{phiprim2} \\
&&\int_{a_{0}}^{a}{\frac{1}{aH}\sqrt{\frac{%
(2D_{L}-3n^{2}M_{p}^{2}L^{2}-\alpha )\left( 1-\sqrt{\frac{%
3M_{p}^{2}L^{2}\Omega _{\Lambda }}{D_{L}}}\text{cosn}(\sqrt{|k|}y)\right) }{%
D_{L}}-\frac{b^{2}(1+\Omega _{k})}{\Omega _{\Lambda }}}da}.
\nonumber
\end{eqnarray}%
Here $a_{0}$ is the present value of the scale factor. Here, we have
established an interacting entropy-corrected holographic tachyon
dark energy model and reconstructed the potential and the dynamics
of the tachyon field. Setting $\alpha =\beta =0$ and $b=0$ in Eqs.
(\ref{phiprim1}) and (\ref{phiprim2}), one can easily see that in
the case of non-interacting HDE, the evolutionary form of tachyon
field in non-flat universe can be obtained as follows
\begin{equation}
\phi (a)-\phi (a_{0})=\sqrt{\frac{2}{3}}\int_{a_{0}}^{a}\frac{1}{aH}%
\sqrt{1-\frac{\sqrt{\Omega_{\Lambda}}}{n}\text{cosn}(\sqrt{|k|}y)}%
da
\end{equation}

\section{Interacting ECHDE K-essence model\label{sec4}}

The K-essence scalar field model can explain the late time
acceleration of the universe. The general scalar field action for
K-essence
model as a function of $\phi $ and $\chi =\dot{\phi}^{2}/2$ is given by \cite%
{5}
\begin{equation}
S=\int d^{4}x\sqrt{-g}\text{ }p(\phi ,\chi ),
\end{equation}%
where the Lagrangian density $p(\phi ,\chi )$ relates to a pressure density
and energy density through the following equations:
\begin{equation}
p(\phi ,\chi )=f(\phi )(-\chi +\chi ^{2}),
\end{equation}%
\begin{equation}
\rho (\phi ,\chi )=f(\phi )(-\chi +3\chi ^{2})\label{Rok}.
\end{equation}%
Hence, the EoS parameter of K-essence scalar field is obtained as
\begin{equation}
\omega _{K}=\frac{p(\phi ,\chi )}{\rho (\phi ,\chi )}=\frac{\chi -1}{3\chi -1%
}.  \label{w_k}
\end{equation}%
In order to consider the K-essence field as a description of the
interacting ECHDE density, we establish the correspondence between
the K-essence EoS parameter,$w_{K}$, and the interacting ECHDE EoS
parameter, $w_{\Lambda }$. Equating Eq.(\ref{Rok}) with
Eq.(\ref{rhoS}), we have
\begin{equation}
f(\phi)=\frac{D(L)}{L^4}\left[\frac{(1-3w_{\Lambda})^2}{2(1-w_{\Lambda})}\right]
\end{equation}
while by equating Eq.(\ref{w_k}) with Eq.(\ref{eos_ec}), we can find
the expression for $\chi $ as
\begin{eqnarray}
&&\chi =\frac{w_{\Lambda }-1}{3w_{\Lambda }-1}
\label{dil1}=\\
&&\frac{1}{3}\left[\frac{-6+2(2D_{L}-3n^{2}M_{p}^{2}L^{2}-\alpha )\left( 1-\sqrt{\frac{%
3M_{p}^{2}L^{2}\Omega _{\Lambda
}}{D_{L}}}\text{cosn}(\sqrt{|k|}y)\right)
D_{L}^{-1}-3b^{2}(1+\Omega _{k})\Omega _{\Lambda }^{-1}}{%
-4+2(2D_{L}-3n^{2}M_{p}^{2}L^{2}-\alpha )\left( 1-\sqrt{\frac{%
3M_{p}^{2}L^{2}\Omega _{\Lambda
}}{D_{L}}}\text{cosn}(\sqrt{|k|}y)\right) D_{L}^{-1}-3b^{2}(1+\Omega
_{k})\Omega _{\Lambda }^{-1}}\right].  \nonumber \label{chi2}
\end{eqnarray}%
Using $\dot{\phi}^{2}=2\chi $, and $\dot{\phi}=\phi ^{\prime }H$, we get
\begin{eqnarray}
&&\phi ^{\prime } =\frac{1}{H}\sqrt{\frac{2}{3}}\sqrt{1+\frac{2}{1-3w _{\Lambda}}}=\frac{1}{H}\sqrt{\frac{2}{3}}\label{phi_prime3}\times\\
&&\sqrt{\frac{-6+2(2D_{L}-3n^{2}M_{p}^{2}L^{2}-\alpha )\left( 1-\sqrt{\frac{%
3M_{p}^{2}L^{2}\Omega _{\Lambda
}}{D_{L}}}\text{cosn}(\sqrt{|k|}y)\right)
D_{L}^{-1}-3b^{2}(1+\Omega _{k})\Omega _{\Lambda }^{-1}}{%
-4+2(2D_{L}-3n^{2}M_{p}^{2}L^{2}-\alpha )\left( 1-\sqrt{\frac{%
3M_{p}^{2}L^{2}\Omega _{\Lambda
}}{D_{L}}}\text{cosn}(\sqrt{|k|}y)\right) D_{L}^{-1}-3b^{2}(1+\Omega
_{k})\Omega _{\Lambda }^{-1}}}.\nonumber
\end{eqnarray}%
Integrating Eq.(\ref{phi_prime3}), we obtain the evolutionary form
of the K-essence scalar field as
\begin{eqnarray}
&&\phi (a)-\phi (a_{0})
=\sqrt{\frac{2}{3}}\int_{a_{0}}^{a}{\frac{d a}{a H}\label{phi_equation}\times}\\
&&{\sqrt{\frac{-6+2(2D_{L}-3n^{2}M_{p}^{2}L^{2}-\alpha )\left( 1-\sqrt{\frac{%
3M_{p}^{2}L^{2}\Omega _{\Lambda
}}{D_{L}}}\text{cosn}(\sqrt{|k|}y)\right)
D_{L}^{-1}-3b^{2}(1+\Omega _{k})\Omega _{\Lambda }^{-1}}{%
-4+2(2D_{L}-3n^{2}M_{p}^{2}L^{2}-\alpha )\left( 1-\sqrt{\frac{%
3M_{p}^{2}L^{2}\Omega _{\Lambda
}}{D_{L}}}\text{cosn}(\sqrt{|k|}y)\right) D_{L}^{-1}-3b^{2}(1+\Omega
_{k})\Omega _{\Lambda }^{-1}}}}.\nonumber
\end{eqnarray}%
We see from Eq. (\ref{phi_prime3}), the kinetic energy
$\dot{\phi}^2$ may exist provided that
\begin{equation}
w_{\Lambda}\leq 1.
\end{equation}
This condition implies that the phantom divide can be crossed in a
universe with an accelerated expansion.\newline
In the limiting case of $\alpha =\beta =0$ and $b=0$, Eqs. (\ref{chi2}), (\ref%
{phi_prime3}) and (\ref{phi_equation}) reduce to the following
simple form for non-interacting holographic K-essence model in
non-flat universe.
\begin{equation}
\chi =\frac{1}{3}\frac{2+\
\frac{\sqrt{\Omega_{\Lambda}}}{n}\text{cosn}(\sqrt{|k|}y) }{1+
\frac{\sqrt{\Omega_{\Lambda}}}{n}\text{cosn}(\sqrt{|k|}y) }
\end{equation}%
\begin{equation}
\phi ^{\prime }=\sqrt{\frac{2}{3}}\frac{1}{H}\sqrt{\frac{2+\
\frac{\sqrt{\Omega_{\Lambda}}}{n}\text{cosn}(\sqrt{|k|}y) }{1+
\frac{\sqrt{\Omega_{\Lambda}}}{n}\text{cosn}(\sqrt{|k|}y) } }
\end{equation}%
\begin{equation}
\phi (a)-\phi (a_{0})=\sqrt{\frac{2}{3}}\int_{a_{0}}^{a}{\frac{1}{a H}\sqrt{%
\frac{2+\ \frac{\sqrt{\Omega_{\Lambda}}}{n}\text{cosn}(\sqrt{|k|}y) }{1+%
\frac{\sqrt{\Omega_{\Lambda}}}{n}\text{cosn}(\sqrt{|k|}y) } }}d a
\end{equation}

\section{Interacting ECHDE Dilaton model \label{sec5}}

The dilaton scalar field is originated from the lower limit of
string theory \cite{dilaton}. The dilaton filed is described by the
effective Lagrangian density as
\begin{equation}
p_{D}=-\chi +ce^{\lambda \phi }\chi ^{2},
\end{equation}%
where $c$ and $\lambda $ are positive constant. Considering the
dilaton field as a source of the energy-momentum tensor in Einstein
equations, one can find that the Lagrangian density corresponds to
the pressure of the scalar field and the energy density of dilaton
field is also obtained as
\begin{equation}
\rho _{D}=-\chi +3ce^{\lambda \phi }\chi ^{2},  \label{rhod}
\end{equation}%
where $2\chi =\dot{\phi}^{2}$. The negative coefficient of the
kinematic term of the dilaton field in Einstein frame makes a
phantom like behavior for dilaton field. The EoS parameter of
dilaton is given by
\begin{equation}
\omega _{D}=\frac{p_{D}}{\rho _{D}}=\frac{-1+ce^{\lambda \phi }\chi }{%
-1+3ce^{\lambda \phi }\chi }.  \label{dilaton2}
\end{equation}%
In order to consider the dilaton field as a description of the
interacting ECHDE density, we establish the correspondence between
the dilaton EoS parameter,$w_{D}$, and the EoS parameter $w_{\Lambda
}$ of interacting ECHDE.  By equating Eq.(\ref{dilaton2}) with
Eq.(\ref{eos_ec}), we can find
\begin{eqnarray}
&&ce^{\lambda \phi }\chi=\frac{w_{\Lambda }-1}{3w_{\Lambda }-1}
\label{dil1}=\label{dilki}\\
&&\frac{1}{3}\left[\frac{-6+2(2D_{L}-3n^{2}M_{p}^{2}L^{2}-\alpha
)\left( 1-\sqrt{\frac{3M_{p}^{2}L^{2}\Omega _{\Lambda
}}{D_{L}}}\text{cosn}(\sqrt{|k|}y)\right)
D_{L}^{-1}-3b^{2}(1+\Omega _{k})\Omega _{\Lambda }^{-1}}{%
-4+2(2D_{L}-3n^{2}M_{p}^{2}L^{2}-\alpha )\left( 1-\sqrt{\frac{%
3M_{p}^{2}L^{2}\Omega _{\Lambda
}}{D_{L}}}\text{cosn}(\sqrt{|k|}y)\right) D_{L}^{-1}-3b^{2}(1+\Omega
_{k})\Omega _{\Lambda }^{-1}}\right].  \nonumber
\end{eqnarray}%
using $\chi =\dot{\phi}^{2}/2$ and $\dot{\phi}=\phi ^{\prime }/H$, one can
rewrite (\ref{dil1}) with respect to $\phi $ as
\begin{eqnarray}
&&\frac{d}{d\ln a}e^{\lambda \phi /2} =\frac{1}{H}\frac{\lambda
}{\sqrt{6c}}\sqrt{1+\frac{2}{1-3w
_{\Lambda}}}=\frac{1}{H}\frac{\lambda }{\sqrt{6c}}
\times   \label{dil2} \\
&&\sqrt{\frac{-6+2(2D_{L}-3n^{2}M_{p}^{2}L^{2}-\alpha )\left(
1-\sqrt{ \frac{3M_{p}^{2}L^{2}\Omega _{\Lambda
}}{D_{L}}}\text{cosn}(\sqrt{|k|} y)\right)
D_{L}^{-1}-3b^{2}(1+\Omega _{k})\Omega _{\Lambda }^{-1}}{
-4+2(2D_{L}-3n^{2}M_{p}^{2}L^{2}-\alpha )\left( 1-\sqrt{\frac{
3M_{p}^{2}L^{2}\Omega _{\Lambda
}}{D_{L}}}\text{cosn}(\sqrt{|k|}y)\right) D_{L}^{-1}-3b^{2}(1+\Omega
_{k})\Omega _{\Lambda }^{-1}}},  \nonumber
\end{eqnarray}
Exactly the same as Sec. \ref{sec4}, from Eq. (\ref{dilki}), one can
see that the kinetic energy may exist provided that $w_{\Lambda}\leq
1$ which implies that the phantom divide can be crossed in a
universe with an accelerated expansion.\newline
Finally the evolutionary form of the dilaton scalar field is written as%
\begin{eqnarray}
&&\phi (a) =\frac{2}{\lambda }\ln \{e^{\lambda \phi
(a_{0})/2}+\frac{\lambda
}{\sqrt{6c}}\int_{a_{0}}^{a}\frac{d a}{a H}\times   \label{dil3} \\
&&\sqrt{\frac{-6+2(2D_{L}-3n^{2}M_{p}^{2}L^{2}-\alpha )\left( 1-\sqrt{%
\frac{3M_{p}^{2}L^{2}\Omega _{\Lambda }}{D_{L}}}\text{cosn}(\sqrt{|k|}%
y)\right) D_{L}^{-1}-3b^{2}(1+\Omega _{k})\Omega _{\Lambda }^{-1}}{%
-4+2(2D_{L}-3n^{2}M_{p}^{2}L^{2}-\alpha )\left( 1-\sqrt{\frac{%
3M_{p}^{2}L^{2}\Omega _{\Lambda
}}{D_{L}}}\text{cosn}(\sqrt{|k|}y)\right) D_{L}^{-1}-3b^{2}(1+\Omega
_{k})\Omega _{\Lambda }^{-1}}}\}.  \nonumber
\end{eqnarray}%
In the limiting case of $\alpha =\beta =0$ and $b=0$, Eq.(
\ref{dil3}) reduce to the following simple form for non-interacting
holographic dilaton model in non-flat universe. Therefore the
evolutionary form of the dilaton scalar field can be
obtained as%
\begin{equation}
\phi (a)=\frac{2}{\lambda }\ln {\left[e^{\lambda \phi
(a_{0})/2}+\frac{ \lambda }{\sqrt{6c}}\int_{a_{0}}^{a}\frac{1}{a
H}\sqrt{\frac{2+\
\frac{\sqrt{\Omega_{\Lambda}}}{n}\text{cosn}(\sqrt{|k|}y) }{1+
\frac{\sqrt{\Omega_{\Lambda}}}{n}\text{cosn}(\sqrt{|k|}y) }}d
a\right] }.
\end{equation}

\section{Conclusion \label{sec6}}

Including the quantum correction in the entropy-area relation, we
studied the interacting ECHDE model in non-flat FRW universe. As is
well known, the scalar field models are the effective description of
an underlying theory of dark energy. We established the
correspondence between tachyon, K-essence and dilaton scalar field
models with interacting ECHDE model. By this correspondence, the
scalar fields may be reconstructed so that they can describe the
evolutionary behavior of interacting ECHDE model. The reconstruction
of potentials and dynamics of these scalar fields which describe the
tachyon, K-essence and dilaton cosmology was obtained. It has been
shown that the phantom divide can not be crossed in ECHDE tachyon
model while it is achieved for ECHDE K-essence and ECHDE dilaton
scenarios. By putting $\alpha=\beta=0$, the EoS parameter of HDE,
potentials and the evolutionary form of HDE scalar field models have
been calculated.

\end{document}